\documentclass[aps,prb,amsmath,reprint,superscriptaddress,showpacs,showkeys,floatfix]{revtex4-1}

\usepackage{graphics}
\usepackage{dcolumn}

\begin{document}

\title{Far Infrared Slab Lensing and Subwavelength Imaging in Crystal Quartz}

\author{R. Estev\^{a}m da Silva}
\affiliation{Departamento de F\'{\i}sica, Universidade do Estado do
Rio Grande do Norte, Costa e Silva, 59610-090 Mossor\'o RN - Brazil}

\author{R. Mac\^{e}do}
\affiliation{Departamento de F\'{\i}sica, Universidade do Estado do
Rio Grande do Norte, Costa e Silva, 59610-090 Mossor\'o RN - Brazil}

\author{T. Dumelow}
\email[Corresponding author: ]{tdumelow@yahoo.com.br}
\affiliation{Departamento de F\'{\i}sica, Universidade do Estado do
Rio Grande do Norte, Costa e Silva, 59610-090 Mossor\'o RN - Brazil}

\author{J. A. P. da Costa}
\affiliation{Departamento de F\'{\i}sica, Universidade do Estado do
Rio Grande do Norte, Costa e Silva, 59610-090 Mossor\'o RN - Brazil}

\author{S. B. Honorato}

\affiliation{Departamento de F\'{\i}sica, Universidade Federal do
Cear\'a, Campus Pici, 60455-900 Fortaleza CE, Brazil}

\author{A. P. Ayala}

\affiliation{Departamento de F\'{\i}sica, Universidade Federal do
Cear\'a, Campus Pici, 60455-900 Fortaleza CE, Brazil}

\date{\today}

\begin{abstract}
We examine the possibility of using negative refraction stemming
from the phonon polariton response in an anisotropic crystal to create a
simple slab lens with plane parallel sides, and show that imaging
from such a lens should be possible at room temperature despite the effects of
absorption that are inevitably present due to phonon damping. In
particular, we consider the case of crystal quartz, a system for
which experimental measurements consistent with all-angle negative
refraction associated with the phonon polariton response have already been demonstrated. Furthermore, we investigate
the possibility of subwavelength imaging from such materials, and
show that it should be possible for certain configurations.
\end{abstract}


\maketitle


\section{\label{sec:intro}Introduction}

The idea of a slab lens stemming from negative refraction was
described by Veselago as far back as 1968.\cite{veselago68} Such a
lens would have plane parallel sides, and an object placed on
one side of the lens would project a real image within the
slab followed by a second image on the other side of it. At
the time, the concept, based on materials having both permittivity
$\varepsilon$ and permeability $\mu$ simultaneously negative, was
regarded as essentially hypothetical. Pendry's 2000
paper\cite{pendry00} brought it into the limelight, however, partly
due to the realization that double negative materials
($\varepsilon<0,\mu<0$) were becoming a reality through metamaterial
engineering,\cite{pendry99} and partly as a result of the suggestion that the resulting
lenses may have imaging possibilities beyond the traditional
diffraction limit, a phenomenon often described as superlensing. Under ideal conditions, this would correspond to perfect imaging.

Although perfect imaging requires exact material parameters that are difficult (if not impossible) to achieve in practice,\cite{guangxia09,shen09} any slab of material that displays negative refraction (defined
in terms of ray or power flow directions) for both positive and
negative angles of incidence should display some degree of slab
lensing behavior regardless of the mechanism leading to negative
refraction. Thus, assuming the slab is sufficiently thick to create
the intermediate image, a second image should be formed on the other side of the slab for a certain range of incident angles (although this does not necessarily imply that superlensing, nor indeed aberration-free imaging, will occur). One very simple way of achieving the necessary negative refraction is to make
the slab from a nonmagnetic anisotropic medium two of whose
principal axes have dielectric tensor components of opposing signs.\cite{eritsyan78,dumelow93,belov03,dumelow05,dvorak06,alekseyev06,hoffman07,eritsyan10,wang10,dasilva10} Media of this type are often referred to as hyperbolic media, due to the form of the associated wavevector dispersion.
In the correct configuration, they induce negative
refraction at all incident angles $\theta_1$ in the range
$-90^0\le\theta_1\le90^0$, thus making them particularly
promising for the construction of slab lenses.\cite{dumelow05,liu08,yao09,fang09}

In general, slab lenses of this type do not lead to subwavelength imaging. Nevertheless, in a restricted geometry in which both the object and the image are at, or very close to (i.e. at near-field distances from), the slab surfaces, subwavelength imaging is possible using slabs of materials with whose dispersion takes a hyperbolic (or associated) form. In this case, image formation does not specifically depend on negative refraction within the slab, but rather on propagation of a collimated beam that is essentially perpendicular to the surfaces. Subwavelength object details, which in air only exist as evanescent waves, are then passed from one side of the slab to the other as channeled propagating waves,\cite{belov06,webb06,li07,wang08,liu08a} a phenomenon described as canalization by Belov \textit{et al}.\cite{belov05}

There are a number of methods for obtaining suitable hyperbolic media. In the visible region, it is usual to use metals, incorporated into structures such as multilayers\cite{belov06,webb06,salandrino06,scalora07,shi08} or oriented nanowires embedded in a dielectric\cite{liu08,yao09} to ensure the necessary anisotropy. At far-infrared (terahertz) frequencies, a useful approach to obtaining negative dielectric tensor components is through the phonon polariton response, since the dielectric function of a polar medium becomes negative around the optic phonon frequencies. A suitable anisotropic response may be obtained, for instance, through the growth of semiconductor superlattices, whose dielectric tensor components may take on opposing signs around the phonon frequencies,\cite{raj85,dumelow90,dumelow93a,venger99} thus leading to the required behavior.\cite{dumelow93,alekseyev06}  An alternative method considered for incorporating the necessary anisotropy into the phonon polariton response is to use alkali halides in aligned rod structures.\cite{foteinopoulou11,reyescoronado12} However, it should not be forgotten that, around the optic phonon frequencies, the dielectric tensor of certain
natural anisotropic crystals may display the required characteristics, with the associated all-angle negative refraction, without the necessity of growing artificial metamaterial structures.\cite{dumelow05,dvorak06,eritsyan10,wang10} Note that all-angle negative refraction of this type should not be confused with negative refraction due to conventional birefringence in a uniaxial crystal whose surface is cut obliquely to the optic axis.\cite{chen05} Such negative refraction only occurs over a small range of incident angles, and the condition necessary for slab lensing (that negative refraction occurs for both positive and negative angles of incidence) is not satisfied. Around the optic phonon frequencies, however, the dielectric tensor components may take opposing signs, resulting in all-angle negative refraction associated with hyperbolic dispersion, without the need for an oblique cut.

Dumelow \textit{et al}\cite{dumelow05} have considered slab lensing based on this principle in the anisotropic crystal triglycine sulfate (TGS) at $5\mbox{ K}$. In this low temperature case, phonon damping is extremely low, and absorption effects almost negligible in the configuration considered. Simulations confirm that slab lensing should occur, albeit with aberrations, in a slab of TGS at low temperature.\cite{dumelow05} At room temperature we expect absorption effects to be considerably greater, so it is necessary to investigate whether a similar behavior should also take place in anisotropic crystals at higher temperatures.  Recent experimental measurements have yielded results supporting
the existence of all-angle negative refraction arising from the phonon polariton
response in crystal quartz at room temperature,\cite{dasilva10} so it is natural to wish to study how
this material may perform as a slab lens. We consider such behavior in the present paper, as well as investigating the possibility of using natural crystals such as quartz for subwavelength imaging.

The structure of the paper is as follows. In Section \ref{sec:allang} we briefly outline the basis of all-angle negative refraction in dielectric hyperbolic media. In Section \ref{sec:phonons} we discuss how this occurs in natural crystals such as quartz, and show supporting experimental data. In Section \ref{sec:slab} we consider the use of natural crystals as slab lenses based on negative refraction, configured to form an image away from the near-field regime. In this case, the image is diffraction-limited. In Section \ref{sec:subwavelength}, in contrast, we consider a configuration, based on canalization, in which subwavelength imaging should be possible. Discussions of the results and future prospects are presented in Section \ref{sec:discussion} and conclusions in Section \ref{sec:conclusions}.

\section{\label{sec:allang}All-Angle Negative Refraction in Anisotropic Dielectric Media}

\begin{figure}
\includegraphics[45,260][260,710]{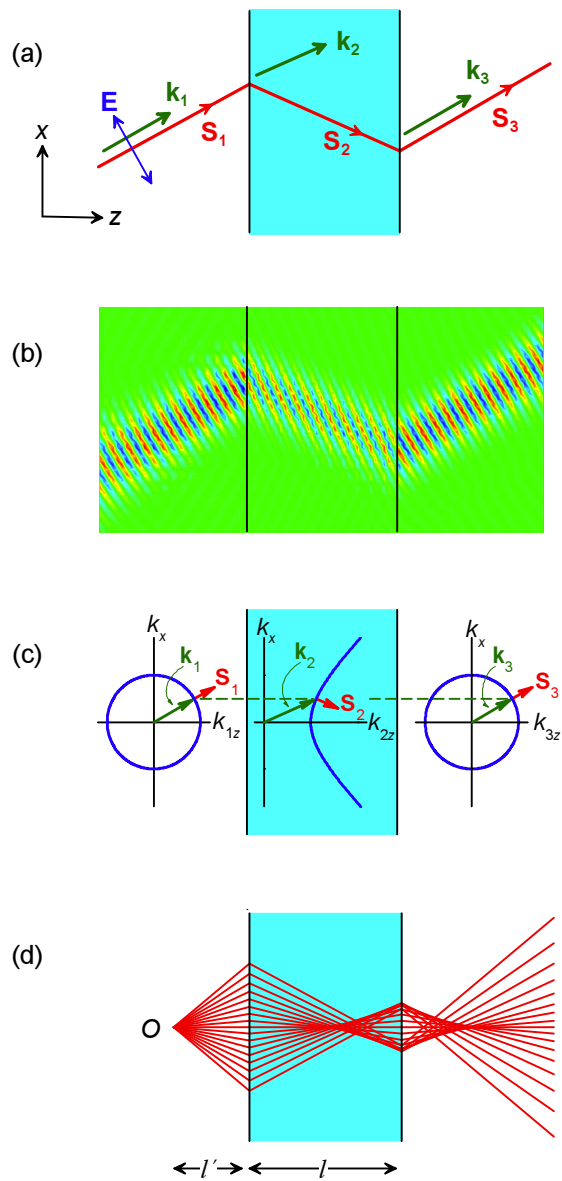}
\caption {\label{ray} (Color online) (a) Wavevector and Poynting vector directions for a p-polarized obliquely incident ray passing through a slab of nonmagnetic anisotropic material. In this example the angle of incidence is $30^0$ and the slab dielectric tensor components are $\varepsilon_{xx}=1$, $\varepsilon_{zz}=-1$. (b) Field profiles showing beam and wavefront directions for a ray passing through the slab.
(c) Equifrequency plots (blue curves) in the three regions, along with the resulting Poynting vector directions normal to the curves. (d) Ray diagram showing the path of several rays passing through the same slab. Ray directions are those of the Poynting vector.}
\end{figure}

In order to understand slab lensing behavior in nonmagnetic anisotropic media, we first consider how negative refraction of a single ray may occur in the geometry shown Fig. \ref{ray}(a). The slab is made of such a medium oriented with its principal axes along the cartesian axes $x$, $y$ and $z$. $xz$ is the plane of incidence (i.e. $k_y=0$) and $z$ is normal to the slab surface. We consider the incident radiation to be $p$-polarized (\textbf{E} field in the plane of incidence $xz$).

If the angle of incidence is represented as $\theta_1$, the in-plane wavevector component $k_x$ is given by
\begin{equation}
\label{kx} k_x=k_0 \sin \theta_1
\end{equation}
where $k_0=\omega/c$. The $z$ component of the
wavevector outside the slab is given by
\begin{equation}
\label{qz1p}
k_{1z}^2 = k_{3z}^2 = k_0^2 - k_x^2
\end{equation}
where the subscripts 1 and 3 represent regions to the left and to the right of the slab respectively.
Inside the slab $k_{2z}$ is represented by
\begin{equation}
\label{qz2p} k_{2z}^2 = k_0^2 \varepsilon_{xx} - k_x^2
\frac{\varepsilon_{xx}}{\varepsilon_{zz}},
\end{equation}
where $\varepsilon_{xx}$ and $\varepsilon_{zz}$ represent two of the
principal components of the dielectric function of the anisotropic
medium. The correct sign of $k_{2z}$ is determined by the condition
that power flow must be away from the interface.\cite{dumelow93}

In defining the angle of refraction, one should remember that ray directions follow the Poynting vector $\mathbf{S}$ rather than the wavevector $\mathbf{k}$. In addition, it is the Poynting vector direction that determines the focusing behavior. We therefore consider the angle of refraction $\theta_2$ in terms of the direction of the Poynting vector $\mathbf{S}_2$ within the slab. Its time-averaged value is given
by $\langle\begin{bf}S\end{bf}_2\rangle = 1/2\mbox{
Re}\left(\begin{bf}E\end{bf} \times
\begin{bf}H^\ast\end{bf}\right)$, leading to a $\theta_2$ value obtained from the expression\cite{dumelow05}
\begin{equation}
\label{angle} \tan{\theta_2}=\frac{\langle S_{2x}\rangle}{\langle
S_{2z}\rangle}=\frac{\mbox{Re}(k_x/\varepsilon_{zz})}{\mbox{Re}(k_{2z}/\varepsilon_{xx})}.
\end{equation}

It is immediately seen that, in general, the direction of the Poynting vector direction $\mathbf{S}_2$ is different from that of the wavevector $\mathbf{k}_2$. This behavior only occurs in $p$-polarization, and only if $\varepsilon_{xx}$ and $\varepsilon_{zz}$ are different. In order to gain a physical understanding of how this may lead to negative refraction, we can, to a first approximation, ignore absorption, leaving both $\varepsilon_{xx}$ and
$\varepsilon_{zz}$ real. In this case $k_{2z}$ can be either real or
imaginary, propagation into the slab occurring when it is real. We note in particular that this is the case when $\varepsilon_{xx}>0$, $\varepsilon_{zz}<0$, and it is straightforward to show that $k_{2z}$ is positive under these conditions.\cite{dumelow93} A simple comparison of Eqs. (\ref{kx}) and (\ref{angle}) thus shows that
$\theta_1$ and $\theta_2$ have opposing signs, leading to negative refraction, as defined in terms of the power flow directions, within the slab. A ray through the slab thus follows the Poynting vector directions shown in Fig. \ref{ray}(a), with the wave behavior shown in Fig. \ref{ray}(b). Comparison of these two figures also confirms that the wavevectors remain normal to the wavefronts in all layers.

An alternative way of interpreting negative refraction in this type of system is shown in Fig. \ref{ray}(c). Since group velocity is given by $\mathbf{v}_g=\mathbf{\nabla}_k\omega$, the group velocity direction, and hence the Poynting vector direction, should be perpendicular to an equifrequency surface in $k$-space. In the $xz$ plane, the equifrequency surface becomes an equifrequency curve, i.e. a plot of $k_z$ against $k_x$ at a given frequency. Such curves are shown in Fig. \ref{ray}(c)in the three regions, being hyperbolic within the slab and circular in the surrounding air. Since $k_x$ is the same in the three regions and simply obtained from the angle of incidence [Eq. (\ref{kx})], we can determine the Poynting vector direction (perpendicular to the equifrequency curve) in each of them, as shown in the figure. The directions are in agreement with those shown in Figs. \ref{ray}(a) and \ref{ray}(b). Negative refraction of the Poynting vector direction is clearly seen, and it is obvious that such behavior will occur for a both positive and negative incident angles (positive and negative $k_x$).

In the case of a series of rays emanating from an object $O$, the simulation in Fig. \ref{ray}(d) shows image formation both within the slab and at the right hand side of it, although there are aberrations associated with the higher incident angles.\cite{dumelow05,liu08,fang09}

\section{\label{sec:phonons}Negative Refraction due to Phonons in Natural Crystals}

One way of satisfying the condition
$\varepsilon_{xx}>0$, $\varepsilon_{zz}<0$  is to make use of the phonon response in natural anisotropic crystals.\cite{dumelow05,dvorak06,eritsyan10,wang10,dasilva10} If we take a the case of a uniaxial crystal, we can write the dielectric tensor, expressed in relation to the crystal axes, as
\begin{equation}
\label{tensor}
 \mathbf{\varepsilon}=\left( \begin{array}{*{3}c}
 \varepsilon_{\textrm{ord}} & 0 & 0 \\
 0 & \varepsilon_{\textrm{ord}} & 0 \\
 0 & 0 & \varepsilon_{\textrm{ext}} \end{array} \right)\,.
\end{equation}
Here $\varepsilon_{\textrm{ext}}$ refers to the response along the extraordinary axis (the crystal´s uniaxis) and $\varepsilon_{\textrm{ord}}$ to the response along the ordinary axes. Around the phonon frequencies, these components may be written in the form\cite{gervais74}
\begin{equation}
\label{phononsord} \varepsilon_{\textrm{ord}}=\varepsilon_{\infty,\textrm{ord}}\prod_n
\frac{\omega_{\textrm{L}n,\textrm{ord}}^2 - \omega^2 - i \omega
\gamma_{\textrm{L}n,\textrm{ord}}}{ \omega_{\textrm{T}n,\textrm{ord}}^2 - \omega^2 - i
\omega \gamma_{\textrm{T}n,\textrm{ord}}}
\end{equation}
\begin{equation}
\label{phononsext} \varepsilon_{\textrm{ext}}=\varepsilon_{\infty,\textrm{ext}}\prod_n
\frac{\omega_{\textrm{L}n,\textrm{ext}}^2 - \omega^2 - i \omega
\gamma_{\textrm{L}n,\textrm{ext}}}{ \omega_{\textrm{T}n,\textrm{ext}}^2 - \omega^2 - i
\omega \gamma_{\textrm{T}n,\textrm{ext}}}\,,
\end{equation}
where $\omega$ is the frequency, $\varepsilon_{\infty,\textrm{ord}}$ and $\varepsilon_{\infty,\textrm{ext}}$ are
the high frequency dielectric constants, $\omega_{\textrm{T}n,\textrm{ord}}$ and $\omega_{\textrm{T}n,\textrm{ext}}$ are the frequencies of transverse optical (TO) phonons, $\omega_{\textrm{L}n,\textrm{ord}}$ and $\omega_{\textrm{L}n,\textrm{ext}}$  are the frequencies of the longitudinal optical (LO) phonons, and $\gamma_{\textrm{T}n,\textrm{ord}}$, $\gamma_{\textrm{T}n,\textrm{ext}}$, $\gamma_{\textrm{L}n,\textrm{ord}}$ and $\gamma_{\textrm{L}n,\textrm{ext}}$  are the appropriate damping parameters responsible for absorption around the phonon frequencies. Since the phonons polarized along the ordinary and extraordinary axes are inherently different, the various phonon parameters (including the phonon frequencies) contributing to the corresponding tensor components are also different. This raises the possibility of tensor components having opposing signs. In principle, in the absence of damping, this should occur in any polar uniaxial crystal within certain frequencies ranges.\cite{mills74} In practice, however, reasonably strong resonances, with frequencies along the different principal crystal axes well separated in relation to the magnitude of the damping parameters, are needed in order to give useful results.

\begin{figure}
\includegraphics*[0,0][215,270]{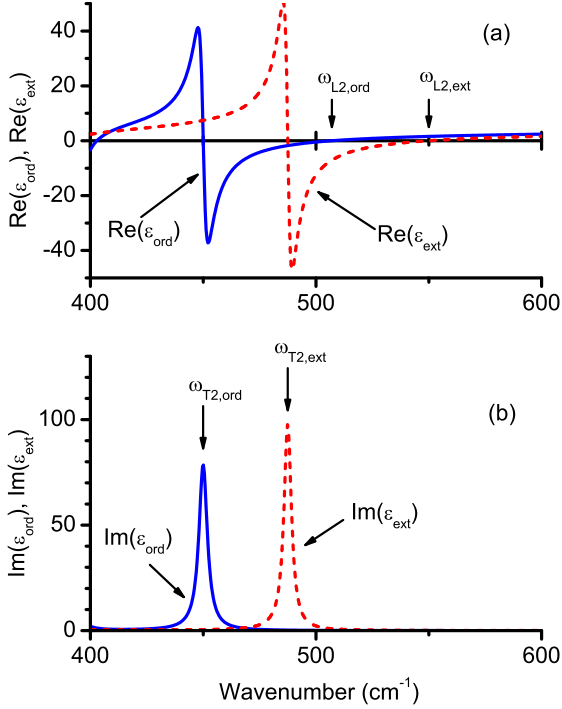}
\caption {\label{diel} (Color online) (a) Real and (b) imaginary parts of the principal components of the dielectric function of quartz in the frequency range \mbox{400 cm}$^{-1}$ to \mbox{600 cm}$^{-1}$.}
\end{figure}

In this paper we consider the case of crystal quartz, a material which shows suitably separated resonances at room temperature.\cite{gervais75,duarte87,wyncke97,brehat97,dasilva10}
In Fig. \ref{diel} we show the values of $\varepsilon_{\textrm{ord}}$ and $\varepsilon_{\textrm{ext}}$ for crystal quartz in the range \mbox{400 cm}$^{-1}$ to \mbox{600 cm}$^{-1}$. The parameters used are based on those obtained by Gervais and Piriou.\cite{gervais75}  We have made adjustments to some of their values, however, in order to give a better fit to the experimental results presented later in this paper, as summarized in Table \ref{table1}.

\begin{table*}
\caption{\label{table1}Comparison of phonon parameters of crystal quartz used in this work with those of Gervais and Piriou.\cite{gervais75} We have retained the high frequency dielectric constants used by these authors ($\varepsilon_{\infty,\textrm{ord}}=2.356$, $\varepsilon_{\infty,\textrm{ext}}=2.383$).}
\begin{ruledtabular}
\begin{tabular}{ccdddddddd}
 &&\multicolumn{4}{c}{Gervais and Piriou}&\multicolumn{4}{c}{This work}\\
 &&\multicolumn{1}{c}{$\omega_{\textrm{T}n}$}
 &\multicolumn{1}{c}{$\gamma_{\textrm{T}n}$}
 &\multicolumn{1}{c}{$\omega_{\textrm{L}n}$}
 &\multicolumn{1}{c}{$\gamma_{\textrm{L}n}$}
 &\multicolumn{1}{c}{$\omega_{\textrm{T}n}$}
 &\multicolumn{1}{c}{$\gamma_{\textrm{T}n}$}
 &\multicolumn{1}{c}{$\omega_{\textrm{L}n}$}
 &\multicolumn{1}{c}{$\gamma_{\textrm{L}n}$}
 \\
 Symmetry&$n$
 &\multicolumn{1}{c}{(cm$^{-1}$)}
 &\multicolumn{1}{c}{(cm$^{-1}$)}
 &\multicolumn{1}{c}{(cm$^{-1}$)}
 &\multicolumn{1}{c}{(cm$^{-1}$)}
 &\multicolumn{1}{c}{(cm$^{-1}$)}
 &\multicolumn{1}{c}{(cm$^{-1}$)}
 &\multicolumn{1}{c}{(cm$^{-1}$)}
 &\multicolumn{1}{c}{(cm$^{-1}$)}

 \\ \hline
 &1&393.5&2.8&402.0&2.8&393.5&2.1&403.0&2.8\\
 &2&450.0&4.5&510.0&4.1&450.0&4.5&507.0&3.5\\
 $E$&3&695.0&13.0&697.6&13.0&695.0\footnotemark[1]&13.0\footnotemark[1]&697.6\footnotemark[1]&13.0\footnotemark[1]\\
 (ordinary)&4&797.0&6.9&810.0&6.9&797.0\footnotemark[1]&6.9\footnotemark[1]&810.0\footnotemark[1]&6.9\footnotemark[1]\\
 &5&1065.0&7.2&1226.0&12.5&1065.0\footnotemark[1]&7.2\footnotemark[1]&1226.0\footnotemark[1]&12.5\footnotemark[1]\\
 &6&1158.0&9.3&1155.0&9.3&1158.0\footnotemark[1]&9.3\footnotemark[1]&1155.0\footnotemark[1]&9.3\footnotemark[1]\\
\hline
&$1$&363.5&4.8&386.7&4.8&363.5\footnotemark[1]&4.8\footnotemark[1]&386.7&7.0\\
$A_2$&2&496.0&5.2&551.5&5.8&487.5&4.0&550.0&3.2\\
(extraordinary)&3&777.0&6.7&790.0&6.7&777.0\footnotemark[1]&6.7\footnotemark[1]&790.0\footnotemark[1]&6.7\footnotemark[1]\\
&4&1071.0&6.8&1229.0&12.0&1071.0\footnotemark[1]&6.8\footnotemark[1]&1229.0\footnotemark[1]&12.0\footnotemark[1]\\
&-&509.0\footnotemark[2]&14.0\footnotemark[2]&507.5\footnotemark[2]&14.0\footnotemark[2]&-&-&-&-\\
\end{tabular}
\end{ruledtabular}
\footnotetext[1]{The parameters associated with modes having frequencies outside the range of this study, or its immediate vicinity, have been left unchanged from those published by Gervais and Piriou.}
\footnotetext[2]{This additional mode, used in modeling the experimental results of Gervais and Piriou, is generally considered to be an experimental artefact,\cite{gervais75,duarte87} and has not been included in our simulations.}\end{table*}

The dielectric function in the phonon region is, in general, complex, but it is reasonable, as a first approximation, to simply look at the its real part in considering the refracting behavior. Thus, for negative refraction to take place, $\mbox{Re}(\varepsilon_{\textrm{ord}})$ and $\mbox{Re}(\varepsilon_{\textrm{ext}})$ should have opposing signs. It is seen that, in the case of crystal quartz, $\mbox{Re}(\varepsilon_{\textrm{ord}})>0$, $\mbox{Re}(\varepsilon_{\textrm{ext}})<0$ in the frequency region between $\omega_{\textrm{L}2,\textrm{ord}}$ and $\omega_{\textrm{L}2,\textrm{ext}}$ (using the mode numbering of Table \ref{table1}) whereas $\mbox{Re}(\varepsilon_{\textrm{ord}})<0$, $\mbox{Re}(\varepsilon_{\textrm{ext}})>0$ in the frequency region between $\omega_{\textrm{T}2,\textrm{ord}}$ and $\omega_{\textrm{T}2,\textrm{ext}}$. The corresponding region of negative refraction depends on the crystal orientation.

\begin{figure}
\includegraphics*[0,80][215,545]{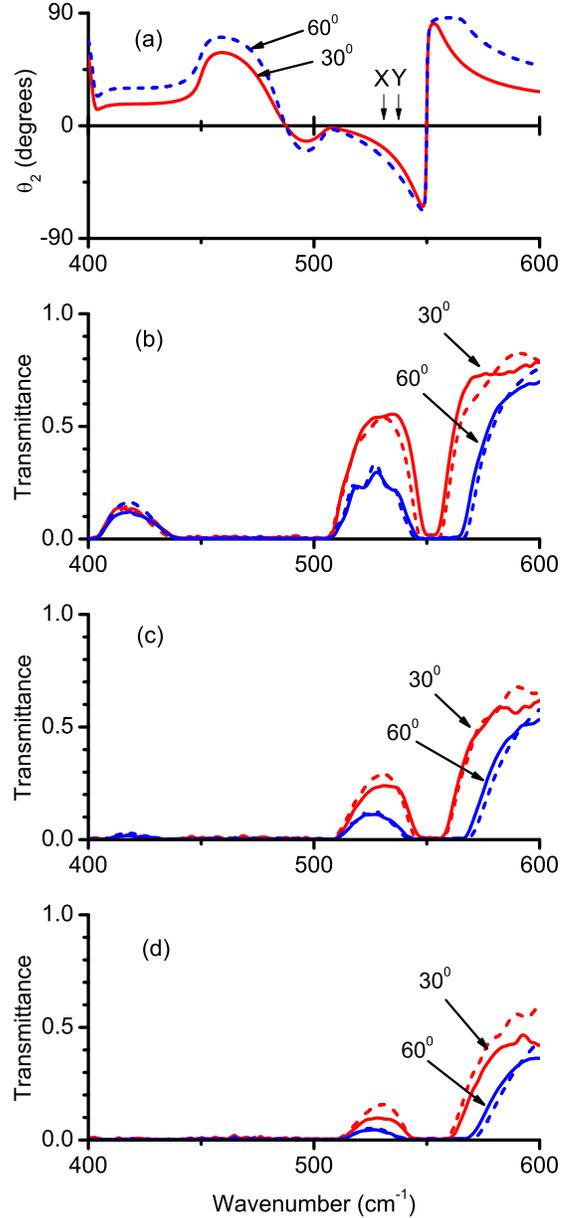}
\caption {\label{transmissionz} (Color online) (a) Simulated p-polarization angle of refraction in the case of quartz oriented with its extraordinary axis along $z$, for incident angles $\theta_1$ of $30^0$ and $60^0$; (b) transmission spectra through a crystal of thickness $l=25\mbox{ }\mu$m at these angles of incidence; (c) transmission spectra through a crystal with $l=50\mbox{ }\mu$m; (d) transmission spectra through a crystal with $l=75\mbox{ }\mu$m. The solid and dashed lines in the transmission spectra represent experimental and simulated results respectively.}
\end{figure}

We first consider the extraordinary axis to be along $z$ (i.e. normal to the crystal surface), so that $\varepsilon_{xx}= \varepsilon_{\textrm{ord}}$ and $\varepsilon_{zz}=\varepsilon_{\textrm{ext}}$. In this case we have the negatively refracting condition $\mbox{Re}(\varepsilon_{xx})>0$, $\mbox{Re}(\varepsilon_{zz})<0$ between $507\mbox{ cm}^{-1}$ and 550$\mbox{ cm}^{-1}$ (i.e. between $\omega_{\textrm{L}2,\textrm{ord}}$ and $\omega_{\textrm{L}2,\textrm{ext}}$), as shown in Fig. \ref{transmissionz}(a). Negative refraction in this orientation has been studied in Ref. \onlinecite{dasilva10}, which shows that significant transmission occurs in the corresponding frequency region. This is confirmed in transmission spectra shown in Fig. \ref{transmissionz}, which shows both experimental data, measured using a Bruker Vertex 70 spectrometer, and theoretical simulations, obtained using standard transfer matrix techniques,\cite{dumelow93a} for various samples thicknesses $l$ (see Fig. \ref{ray}) and incident angles $\theta_1$.

In order to quantify the efficiency, we can look at the figure of merit. This parameter, often used to characterize negatively refracting media, is traditionally defined as $|\mathrm{Re}(n)|/\mathrm{Im}(n)$, where $n$ represents the refractive index of the material.\cite{shalaev07} In the present case, we follow the lead of Hoffman \emph{et al}\cite{hoffman07} and interpret the figure of merit as $\mathrm{Re}(k_{2z})/\mathrm{Im}(k_{2z})$. At the frequency  marked as $X$ in Fig. \ref{transmissionz} (531$\mbox{ cm}^{-1}$), where the transmission is relatively high, this gives a figure of merit of 31 at $\theta_1=30^0$ and of 23 at $\theta_1=60^0$. These values are considerably larger than those typically encountered for metamaterial structures.

We model the behavior of a finite beam passing through the slabs at this frequency by considering the incident beam to be gaussian, and represent it as a Fourier sum of plane waves:
\begin{equation}
\label{fourier} H_y=\int_{-\infty}^\infty \psi(k_x) e^{i(k_x x +
k_{1z} z)} dk_x.
\end{equation}
In the case of a gaussian beam, $\psi(k_x)$ can be written\cite{horowitz71}
\begin{equation}
\label{gauss} \psi(k_x) = -\frac{g}{2 \cos\theta_0\sqrt\pi} \exp
\left[-\frac{g^2 \left(k_x-k_0 \sin
\theta_0\right)^2}{4\cos^2\theta_0} \right],
\end{equation}
where $2 g$ represents the beam width at its waist and $\theta_0$
represents the effective incident angle of the overall beam. In practice, we assume that all components of the gaussian beam are propagating in air (i.e. $k_{1z}$ is real),\cite{chen04} so we restrict the integral in Eq. (\ref{fourier}) to the range $-k_0\leq k_x \leq k_0$.

Using an incident beam of this form, it is possible to use standard multilayer optics techniques to calculate the magnetic field associated with each plane wave component at any point in the $xz$ plane. Numerical integration then gives the overall \textbf{H} fields, and thus the associated \textbf{E} fields and Poynting vectors.\cite{kong02,dumelow05}

\begin{figure}
\includegraphics*[0,75][250,210]{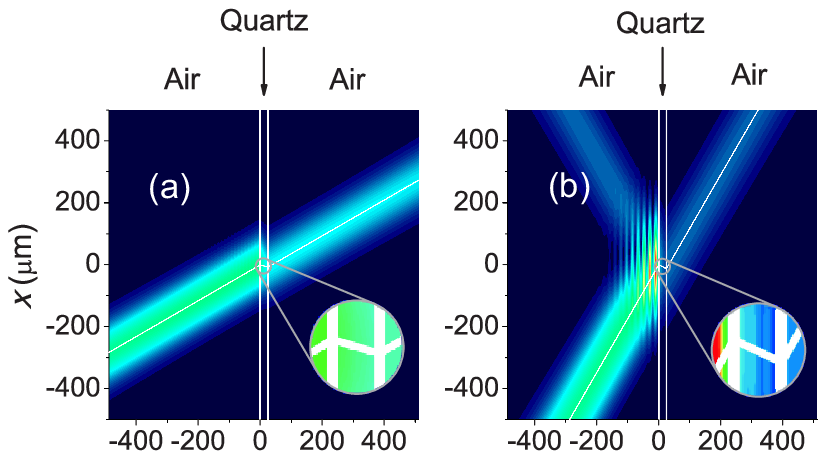}
\includegraphics*[0,75][250,185]{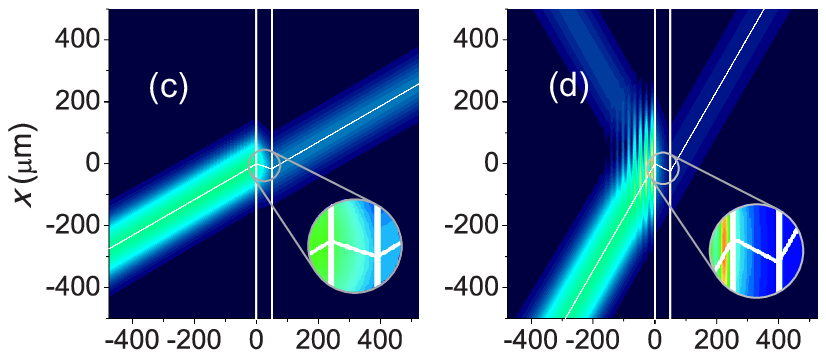}
\includegraphics*[0,60][250,185]{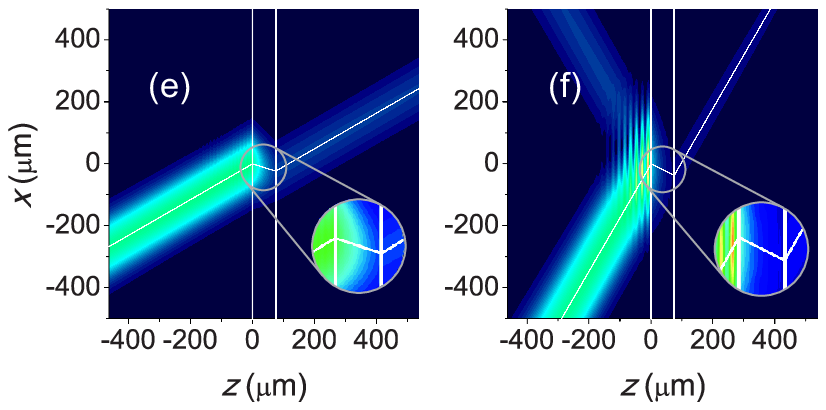}

\caption {\label{gausscontour} (Color online) Simulation of the intensity profile (in terms of the magnitude of the time-averaged Poynting vector) of a gaussian beam passing through a quartz slab in the configurations used in Fig. \ref{transmissionz} at frequency $X$ (531$\mbox{ cm}^{-1}$). The basic geometry is shown in Fig. \ref{ray}(a), with the quartz uniaxis along $z$. (a) $l=25\mbox{ }\mu$m, $\theta_1=30^0$; (b) $l=25\mbox{ }\mu$m, $\theta_1=60^0$; (c) $l=50\mbox{ }\mu$m, $\theta_1=30^0$; (d) $l=50\mbox{ }\mu$m, $\theta_1=60^0$; (e) $l=75\mbox{ }\mu$m, $\theta_1=30^0$; (f) $l=75\mbox{ }\mu$m, $\theta_1=60^0$. The thin white line through the center of the beam represents the ray path calculated using the angle of refraction given by Eq. (\ref{angle}). The insets show details of negative refraction within the slab. Note that the sample thicknesses of $25\mbox{ }\mu$m, $50\mbox{ }\mu$m and $75\mbox{ }\mu$m correspond to 1.3, 2.7 and 4.0 free-space wavelengths respectively.}
\end{figure}

The resulting beam profiles for the various experimental configurations represented in Fig. \ref{transmissionz} are shown in Fig. \ref{gausscontour}. Here the incident beam, whose width is given by $g=100\mbox{ }\mu$m, is assumed to be focused at the slab surface, at $x=0$, $z=0$. Negative refraction, seen as a displacement of the transmitted beam in the negative $x$ direction in a manner similar to that shown in Fig. \ref{ray}(b), occurs in each case. The displacement is naturally greater for thicker samples, but the transmission is lower, in line with the spectra shown in Fig. \ref{transmissionz}. In addition, the transmitted intensity is significantly reduced when the angle of incidence is increased. This is also observed in the experimental results.

\section{\label{sec:slab}Slab Lensing in Natural Crystals}

We now consider how negative refraction in natural crystals such as quartz may be used for slab lensing of the type shown in Fig. \ref{ray}(d). Rather than the frequency $X$ used in the simulation of transmission of gaussian beams, we find it convenient to show results for the slightly higher frequency marked as $Y$ in Fig. \ref{transmissionz}(a), at 537$\mbox{ cm}^{-1}$. This is because, although the transmission is lower at this frequency (the figure of merit is 28 at $\theta_1=30^0$ and 18 at  $\theta_1=60^0$), the angle of refraction $\theta_2$ is (in magnitude) somewhat higher.

\begin{figure}
\includegraphics*[0,0][215,260]{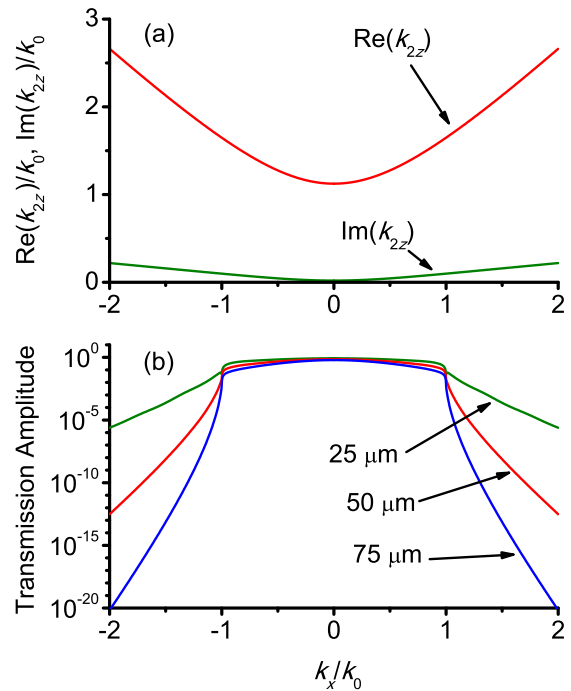}
\caption {\label{transferz} (Color online) (a) Real and imaginary parts of the wavevector component $k_{2z}$ as a function of $k_x$ (expressed in units of $k_0$), for transmission in a quartz crystal having its extraordinary axis directed along $z$, in p-polarization, at frequency $Y$ (537$\mbox{ cm}^{-1}$). (b) Amplitude of the $H_y$ field at the image plane in the configuration shown in Fig. \ref{slitcontour}. Here we take the image position be at the appropriate intensity maximum in Fig. \ref{slitcontour}, which is at $z=45\mbox{ }\mu$m in the case of slab thickness $l=25\mbox{ }\mu$m, $z=94\mbox{ }\mu$m in the case of $l=50\mbox{ }\mu$m, and $z=146\mbox{ }\mu$m in the case of $l=75\mbox{ }\mu$m.}
\end{figure}

Plots of the real and imaginary parts of $k_{2z}$ as a function of $k_x$ (both wavevector components being normalized with respect to $k_0$) at frequency $Y$ are shown in Fig. \ref{transferz}(a). The ratio of these two plots gives the figure of merit. Of particular interest in this figure is the $\mbox{Re}(k_{2z})$ curve, since this is in essence an equifrequency plot. Hyperbolic dispersion of the type shown in Fig. \ref{ray}(c) is clearly present, so slab lensing similar to that in Fig. \ref{ray}(d) should be expected.

\begin{figure}
\includegraphics*[0,0][242,338]{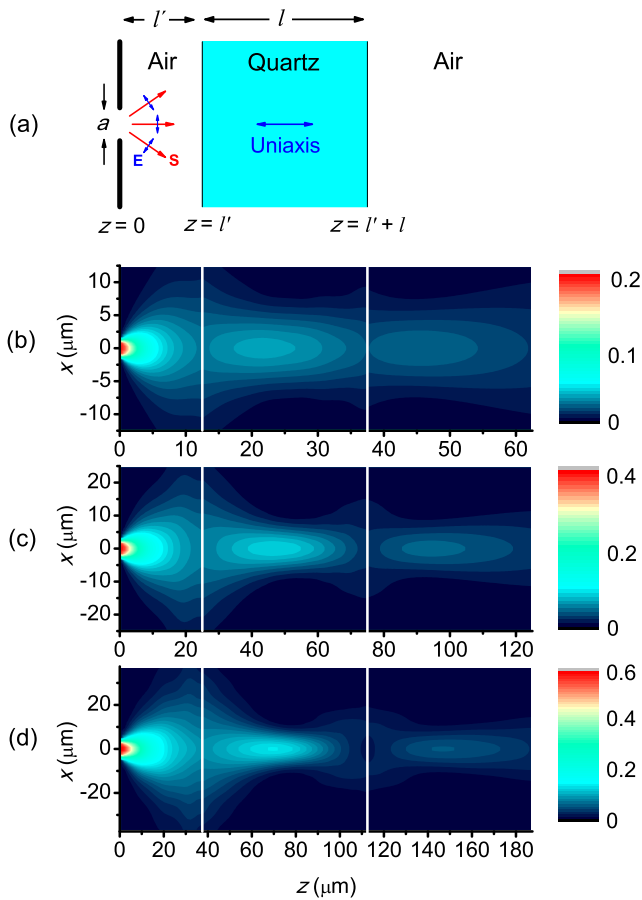}

\caption {\label{slitcontour} (Color online) P-polarization image formation due to a slit source placed to the left of a quartz slab, whose extraordinary axis is directed along $z$ (normal to the slab surface), at frequency $Y$ (537$\mbox{ cm}^{-1}$). (a) Schematic showing the general setup. (b) - (d) Simulation of the intensity profile, using parameters (b) $l=25\mbox{ }\mu$m , $l^\prime=12.5\mbox{ }\mu$m, $a=2.5\mbox{ }\mu$m, (c) $l=50\mbox{ }\mu$m, $l^\prime=25\mbox{ }\mu$m, $a=5\mbox{ }\mu$m, and (d) $l=75\mbox{ }\mu$m, $l^\prime=37.5\mbox{ }\mu$m, $a=7.5\mbox{ }\mu$m. The slit widths $a$ correspond to (b) $0.134\lambda$, (c) $0.267\lambda$, and (d) $0.403\lambda$, where $\lambda$ represents the free-space wavelength.}
\end{figure}

In the slab lens calculations, we take a source to be positioned at $x=0$, $z=0$, at a distance $l^\prime$ to the left of the slab, i.e. the front surface of the slab is at $z=l^\prime$ [see Fig. \ref{slitcontour}(a)]. As an approximation to a slit source, the amplitude of the incident $H_y$ field is assumed constant in the range $-a/2$ to $a/2$ at $z=0$, being zero at all other points in this plane. Thus $a$ effectively represents a slit width. In a manner similar to that used to describe a gaussian beam, we represent the incident field to the right of the $z=0$ plane by Eq.
(\ref{fourier}), but $\psi(k_x)$ is now given by
\begin{equation}
\label{slit} \psi(k_x) = \frac{\sin(k_x a/2)}{\pi k_x}.
\end{equation}
The techniques used for calculating the overall fields and intensities in the $xz$ plane are then the same as those used above for gaussian beam simulations.

The result of intensity simulations in the $xz$ plane are shown in Fig. \ref{slitcontour} for the three slab thicknesses considered earlier. For each of these thicknesses we take $l^\prime=l/2$ and $a=l/10$, so that, if these figures were replaced by ray diagrams (with, for instance, with each side of the slit represented as a single point source) the three figures would be equivalent. In practice, Fig. \ref{slitcontour} shows that focusing of the internal and external images [see Fig. \ref{ray}(d)] occurs in each case, but the image size does not simply scale with the overall dimensions of the system as would occur in a geometric optics analysis.

In interpreting these results, one should note that both object and image are sufficiently far from the slab that near-field effects can be reasonably ignored. Thus evanescent waves from the object play a negligible role in the formation of the image. We can see this from Fig. \ref{transferz}(b), which shows the transfer function, i.e. the amplitude of the transmission coefficient from the object plane to the image plane,\cite{smith03} of each plane wave component in the range $-2 k_0\leq k_x\leq 2 k_0$.  It is seen that the amplitude quickly falls off for $|k_x|>k_0$, the region in which the waves are evanescent in air.  Diffraction-limited imaging should therefore be expected. In interpreting the results of Fig. \ref{slitcontour} with this in mind, we note that the slit width considered in the calculations is smaller than the diffraction limit in each case.  If the image size is diffraction limited, it is therefore natural to expect the most pronounced increase in image size with respect to object size $a$ in the situation shown in Fig. \ref{slitcontour}(b), where the object is smallest, and this is what is indeed observed.

It is clear from the above that, since, in the present setup, subwavelength details associated with $|k_x|>k_0$ tend to get lost, the slabs are not functioning as superlenses of the type considered by Pendry.\cite{pendry00} In the Pendry lens, consisting of a slab with $\varepsilon=-1,\mu=-1$, any decay of the evanescent fields in air is compensated by growing evanescent fields within the slab, so these details are recovered. In the type of medium considered here, however, Fig. \ref{transferz}(a) shows that $k_{2z}$ is essentially real for all $k_x$, i.e. there are no evanescent fields within the slab, either decaying or growing (although the propagating fields may suffer decay due to absorption). Thus, this type of lens can never operate in the same way as a Pendry lens in achieving subwavelength resolution.

Apart from the diffraction-limiting, other effects are important in determining the image quality. Firstly, the effects of absorption are not inconsiderable at the slab thicknesses considered here, as observed in Figs. \ref{transmissionz} and \ref{gausscontour}. This reduces the intensity of the image and, since the effect is larger for larger $k_x$ [see Fig. \ref{transferz}(b)], may also change the intensity distribution of the image. Secondly, even in a geometrical optics analysis, image formation due to slab lenses with $\varepsilon_{xx}>0$, $\varepsilon_{zz}<0$ is not perfect, and there are aberrations of the type shown in Fig. \ref{ray}(d). These aberrations should be more important in thicker slabs such as that shown in Fig. \ref{slitcontour}(d). In thinner slabs of the type shown in Fig. \ref{slitcontour}(b) diffraction-limiting effects somewhat overshadow such aberrations.

\section{\label{sec:subwavelength}Subwavelength Imaging Possibilities}

As discussed above, restoration of evanescent waves is not possible for this type of lens, since evanescent waves are not present within the slab. However, the absence of such evanescent waves may be used to advantage if the slab is placed within the near field of the object. In this case, evanescent waves in air are converted to propagating waves in the slab. At the other side of the slab, these waves may then be converted back to evanescent waves and, given the right slab parameters, contribute to an image with subwavelength resolution at a near-field distance from the slab. In the present section we consider the formation of such subwavelength images, restricting our attention to the extreme case where both the object and the image are actually at the slab surfaces. Thus, in the notation of Figs. \ref{ray}(d) and \ref{slitcontour}(a), we arrange to have $l^\prime=0$ with an image at $z=l$. In this configuration, field attenuation due to evanescent decay in air is reduced to zero.

In order to achieve perfect imaging, the fields associated with each $k_x$ component should all arrive at the image point with the same phase and with the same relative loss of amplitude (although preferably with no loss of amplitude at all). We initially search for a condition that gives a phase change whose dependence on $k_x$ is small. Unless the slab is very thin, the main contribution to the change in phase between object and image will normally be that from transmission within the slab, which depends on the real part of the wavevector component $k_{2z}$. If $\mbox{Re}(k_{2z})$ can be made independent of $k_x$, all components should transmit across within the slab with the same phase, as required. Equation (\ref{qz2p}) shows that this occurs when $\mbox{Re}(\varepsilon_{xx})\geq 0$, $1/\varepsilon_{zz}\rightarrow0$. As seen from Eq. (\ref{angle}), this also corresponds to $\theta_2=0$, so transmission occurs as a collimated beam, which may be of subwavelength width, across the slab.

It is noticeable that the condition $1/\varepsilon_{zz}\rightarrow0$ merely requires that the amplitude of $\varepsilon_{zz}$ be large, without any restriction on its sign. In fact, it is not even required to be real, so a large imaginary $\varepsilon_{zz}$ satisfies the condition. From Fig. \ref{diel}, we see that, within the range investigated, the combined condition $\mbox{Re}(\varepsilon_{xx})\geq 0$, $1/\varepsilon_{zz}\rightarrow0$ does not occur in quartz if the extraordinary axis lies along $z$ ($\varepsilon_{xx}= \varepsilon_{\textrm{ord}}$,  $\varepsilon_{zz}=\varepsilon_{\textrm{ext}}$), but occurs at the TO phonon frequency $\omega_{\textrm{T}2,\textrm{ord}}$ (450$\mbox{ cm}^{-1}$) if the extraordinary axis lies along $x$ ($\varepsilon_{xx}= \varepsilon_{\textrm{ext}}$, $\varepsilon_{zz}=\varepsilon_{\textrm{ord}}$), since $\mbox{Im}(\varepsilon_{zz})$ becomes large at this frequency. This is therefore the geometry and frequency we use in our discussion of subwavelength imaging.

\begin{figure}
\includegraphics*[0,0][230,155]{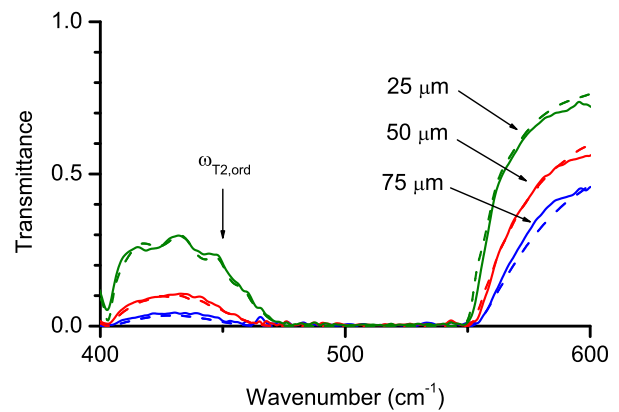}
\caption {\label{transmissionx} (Color online) Oblique incidence p-polarized transmission spectra through various thicknesses of quartz crystals having their extraordinary axes directed along $x$. The angle of incidence $\theta_1$ is $30^0$ in each case. The solid and dashed lines represent experimental and simulated results respectively.}
\end{figure}

Since we are considering transmission across the slab in p-polarization at a resonance frequency, we should first check that there is no absorption associated with this resonance. In Fig. \ref{transmissionx} we show both experimental and theoretical p-polarized transmission spectra in the required geometry at oblique incidence. It is clearly seen that there is no absorption dip at $\omega_{\textrm{T}2,\textrm{ord}}$, even though the $z$-component of the incident \textbf{E} field is nonzero in p-polarization. We can interpret this in the following way. Boundary conditions dictate that $D_z$ should be continuous across the interface, so a large $|\varepsilon_{zz}|$ implies that
$E_z\rightarrow0$ in the slab, and the TO mode is not excited. There is some absorption in this region, as can be observed from the decreasing transmission with increasing slab thickness, but this is mainly due to the proximity of the $x$-polarized TO phonon at $\omega_{\textrm{T}2,\textrm{ext}}$.

\begin{figure}
\includegraphics*[0,0][215,380]{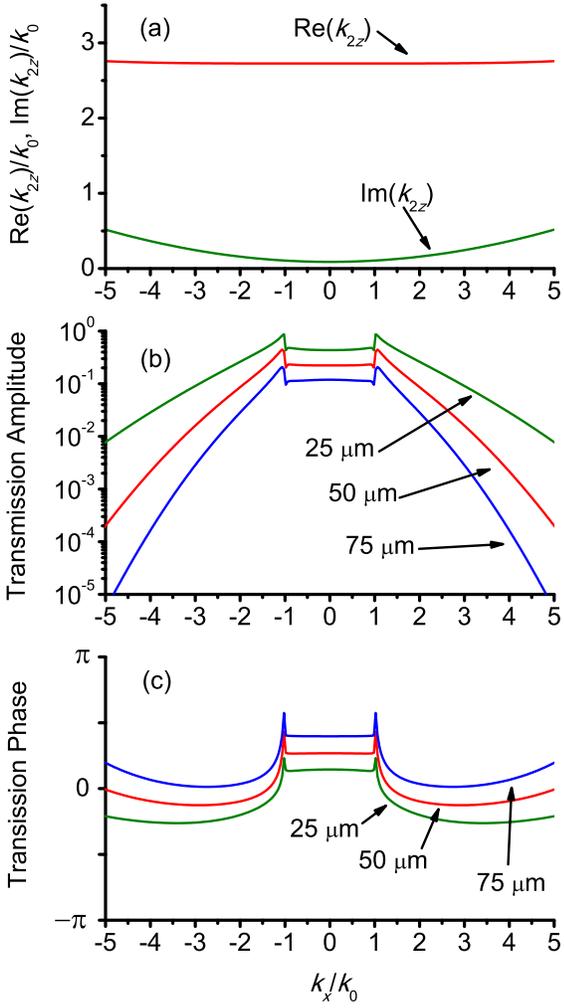}
\caption {\label{transfer} (Color online) (a) Real and imaginary parts of the wavevector component $k_{2z}$ as a function of $k_x$ (expressed in units of $k_0$), for transmission in a quartz crystal having its extraordinary axes directed along $x$, in p´-polarization, at frequency $\omega_{\textrm{T}2,\textrm{ord}}$ (450$\mbox{ cm}^{-1}$). (b) Amplitude and (c) phase of the $H_y$ field transmitted through various thicknesses of crystal quartz in the same configuration.}
\end{figure}

Figure \ref{transfer}(a) shows the real and imaginary parts of $k_{2z}$ as a function of in-plane wavevector $k_x$ (both normalized with respect to $k_0$), and confirms that the condition of nearly constant $\mbox{Re}(k_{2z})$ is satisfied. Thus, if we regard the $\mbox{Re}(k_{2z})$ curve as an equifrequency plot, it is clear that there will be propagation in the $z$ direction for all $k_x$, leading to the required canalization behavior. In addition, we see that $\mbox{Im}(k_{2z})$, which is responsible for absorption, is relatively small. In the region $-k_0<k_x<k_0$, corresponding to real angles of incidence (i.e. propagating waves in air), it is always less than $\mbox{0.03 }\mu \mathrm{m}^{-1}$. This is equivalent to a figure of merit $\mathrm{Re}(k_{2z})/\mathrm{Im}(k_{2z})$ ranging from 31 a normal incidence to 26 at grazing incidence. At higher $|k_x|$, the absorption gradually increases and at $k_x=\pm5k_0$ the figure of merit drops to 5.

Figure \ref{transfer}(b) shows the overall amplitude of the $H_y$ field (i.e. the transfer function) transmitted through slabs of quartz, having the three studied thicknesses, as a function of $k_x/k_0$. Figure \ref{transfer}(c) shows the associated phase. For perfect imaging, both amplitude and phase would be constant for all $k_x$ (the amplitude taking a value equal to unity in the ideal case). In practice, there are noticeable deviations from this behavior.

We can interpret the curves in Fig. \ref{transfer}(b) in terms of two separate effects, transmission efficiency across the two interfaces at either side off the slab and absorption within the slab associated with the imaginary part of $k_{2z}$ shown in Fig. \ref{transfer}(a). The first effect gives the basic shape of the curves and the second effect accounts for the separation of the three curves representing the three different thicknesses and contributes to the drop off in transmission at higher $|k_x|$, where $\mbox{Im}(k_{2z})$ is larger. Figure \ref{transfer}(c) shows that there is some phase change with $k_x$, but the overall variation for a particular slab thickness is around $\pi/2$ in the range shown. Similarly to the amplitude curves discussed above, the basic shape is associated with phase changes on transmission across the interfaces. Phase changes associated with transmission within the slab simply give a vertical shift to this basic shape, since these phase changes are almost independent of $k_x$. From the above, we see that a major restriction to the required subwavelength imaging behavior is likely to be associated with transmission across the interfaces. A number of studies of the use of metallic layered structures to achieve the required anisotropic dielectric tensor, have also discussed this phenomenon.\cite{ramakrishna03,belov06,webb06,li07,wang08,liu08a} Of particular importance is the suggestion that use of a slab thickness equal to an exact number of half-wavelengths (i.e. $\mbox{Re}(k_{2z})l=m\pi$ where $m$ is an integer), thus assuring constructive interference from Fabry-Perot fringes, should overcome these restrictions.\cite{wang08,liu08a} A special case of this, equivalent to choosing $m=0$, is possible if $\varepsilon_{xx}=0$. In the present work, we can see from Fig. \ref{transmissionx} that some weak interference fringes are observed in the transmission spectra in the case of the $25\mbox{ }\mu$m-thick sample, but that they are essentially absent in the case of the thicker samples. Thus we believe that, for the range of thicknesses considered in this work, the Fabry-Perot condition is not of crucial importance since the higher order partial rays are absorbed by the slab. In fact, the $25\mbox{ }\mu$m-thick sample is close to satisfying the Fabry-Perot condition with $m=6$ (an exact calculation gives $m=6.13$), but a small change in the slab thickness does not appear to have much effect on the results.

\begin{figure}
\includegraphics*[0,0][230,405]{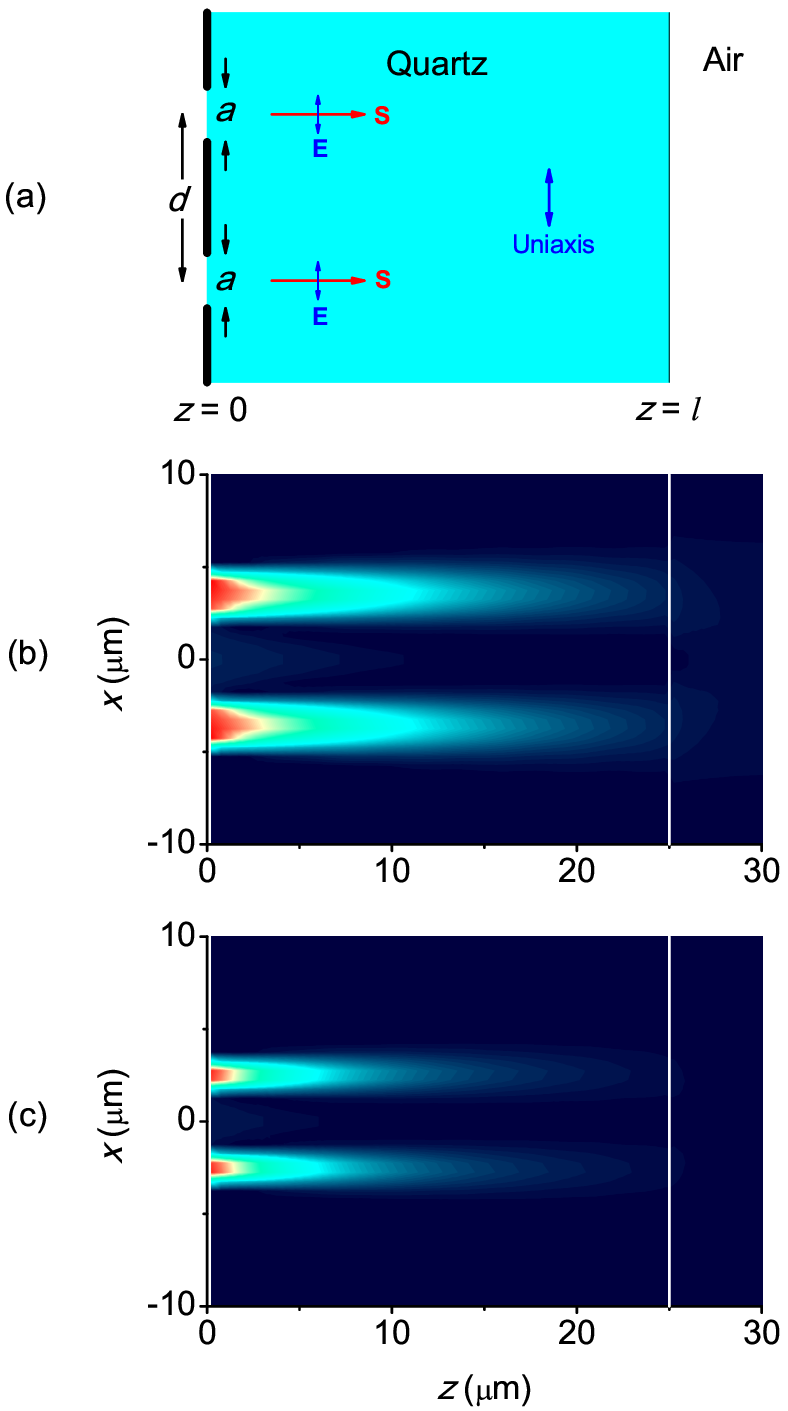}

\caption {\label{2slitcontour} (Color online) Imaging  due to a two-slit source at the surface of a slab of quartz, whose extraordinary axis is along $x$, at frequency $\omega_{\textrm{T}2,\textrm{ord}}$ (450$\mbox{ cm}^{-1}$). (a) Schematic showing the general setup. (b) and (c) Simulation of the intensity profile, using parameters (b) $a=2.5\mbox{ }\mu$m, $d=7\mbox{ }\mu$m and (c) $a=1.5\mbox{ }\mu$m, $d=5\mbox{ }\mu$m. The slab thickness $l$ is $25\mbox{ }\mu$m in each case.}
\end{figure}

We now turn to simulations of subwavelength imaging itself. We consider a two-slit source in which the magnetic field of the incident beam is assumed to be constant across the width of each slit, as before. For slits of width $a$ separated by a distance $d$ [see Fig. \ref{2slitcontour}(a)], this amounts to setting $\psi(k_x)$ to
\begin{equation}
\label{2slit} \psi(k_x) = \frac{2 \sin(k_x a/2)\cos(k_x d/2)}{\pi k_x}.
\end{equation}

\begin{figure}
\includegraphics*[0,45][215,410]{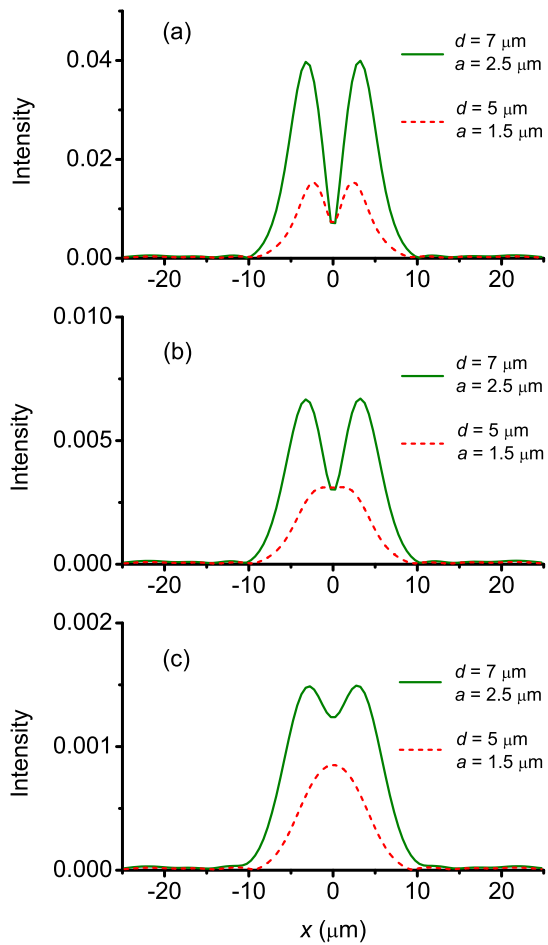}

\caption {\label{2slitprofile} (Color online) Intensity profile transmitted across a slab of quartz with its extraordinary axis along $x$, due to a two-slit source of slit width $a$ and separation $d$. Calculations were made at frequency $\omega_{\textrm{T}2,\textrm{ord}}$ (450$\mbox{ cm}^{-1}$). The intensity scale is normalized with respect to a plane wave whose magnetic field amplitude is that of the incident field in the slits. (a) Slab thickness $l = 25\mbox{ }\mu$m, (b) $l = 50\mbox{ }\mu$m, (c) $l = 75\mbox{ }\mu$m.}
\end{figure}

The slits are placed at the front surface of the slab ($l^\prime=0$). Figures \ref{2slitcontour}(b) and \ref{2slitcontour}(c) show the resulting intensity distributions in the case of the $25\mbox{ }\mu$m-thick slab. Figure \ref{2slitcontour}(b) shows results for a slit separation of $d=7\mbox{ }\mu$m ($0.32\mbox{ }\lambda$) and slit widths $a=2.5\mbox{ }\mu$m ($0.11\mbox{ }\lambda$). The intensities from the two slits are well resolved within the slab, with significant loss of intensity with propagation through the slab. The solid (green) curve in Fig. \ref{2slitprofile}(a) shows the intensity distribution passed through the slab. The images from the two slits are still easily resolved. When the slit separation $d$ is reduced to $d=5\mbox{ }\mu$m ($0.23\mbox{ }\lambda$), we have found that the images are better resolved if we also reduce the slit widths. We therefore show  the intensity distribution for a slit separation of $d=5\mbox{ }\mu$m  and slit widths $a=1.5\mbox{ }\mu$m ($0.07\mbox{ }\lambda$) in Fig. \ref{2slitcontour}(c). The intensities from the two slits are still well resolved within the slab, but are considerably reduced due to the narrowing of the slits. The intensity distribution passed through the slab is shown as the dashed (red) curve in Fig. \ref{2slitprofile})(a), and the two peaks are once more resolved.

Figures \ref{2slitprofile}(b) and \ref{2slitprofile}(c) show the intensity distributions passed through slabs of thicknesses $l=50\mbox{ }\mu$m and $l=75\mbox{ }\mu$m respectively, using the same slit width/separation combinations as for the thinner slab. It is seen that when $l=50\mbox{ }\mu$m the images of the slits are still resolvable for $d=7\mbox{ }\mu$m, but not for $d=5\mbox{ }\mu$m. When $l = 75\mbox{ }\mu$m, some structure remains in the $d=7\mbox{ }\mu$m case, but not in the case of $d=5\mbox{ }\mu$m.

We thus see that subwavelength imaging should occur even for relatively thick slabs of quartz, corresponding to a few free-space wavelengths, albeit with considerable loss of intensity.

\section{\label{sec:discussion}Discussion and Outlook}

The above results confirm that simple anisotropic crystals, such as quartz, should function as slab lenses as well as achieving images with subwavelength resolution. We have restricted our simulations to the slab thicknesses used in our experimental spectral investigations, and such thicknesses are easily obtainable (the quartz crystals used in the spectroscopic measurements were obtained commercially from Boston Piezo-Optics).

The slabs used for our subwavelength studies are fairly thick in relation to those in most studies based on multilayer structures.\cite{belov06,webb06,li07,wang08,liu08a} Better resolution should be possible with thinner slabs, and it should also be possible to take advantage of Fabry-Perot interference in such cases.\cite{wang08,liu08a} Nevertheless, in practice, the behavior at the interfaces may be drastically affected by the source and detector configuration if they are close to the surfaces, and a plane-wave analysis, although correct within the slab, may not give an accurate indication of the interface behavior.

In this study we have only considered crystal quartz as the slab medium, but there are a number of anisotropic crystals that may be suitable. Amongst those considered in the context of hyperbolic behavior are TGS,\cite{dumelow05} Hg$_2$I$_2$,\cite{dvorak06} MgF$_2$,\cite{eritsyan10} and sapphire.\cite{wang10}  Several factors may be important in choosing suitable materials. Obviously different materials will be appropriate for different frequency ranges, and phonon resonances must be sufficiently strong and well separated. Absorption clearly plays a vital role in the image formation, so low damping is important. For subwavelength imaging using the  $\mbox{Re}(\varepsilon_{xx})\geq 0$, $1/\varepsilon_{zz}\rightarrow0$ criterion at $z$-polarized TO frequencies, it would be useful to have negligible Im$(\varepsilon_{xx})$, so there should ideally be no $x$-polarized phonons close to the frequency of interest. Dvorak and Kuzel\cite{dvorak06} discuss the case of Hg$_2$I$_2$ in the context of negative refraction (rather than imaging behavior). The damping parameters for this material are somewhat larger than those for quartz, but its optical phonon frequencies along the principal axes are well separated from one another, so may be suited to subwavelength imaging applications. As mentioned in the introduction, Dumelow \emph{et al}\cite{dumelow05} consider slab lenses from triglycine sulfate (TGS), which, at low temperature, has both very low damping and well-separated phonon frequencies. In addition to the slab lensing properties discussed in the paper, this material is likely to give very good subwavelength imaging. The disadvantage is, of course, the necessity of low temperature. Another material considered as a hyperbolic medium is MgF$_2$, which has properties somewhat similar to quartz.\cite{eritsyan10}

In this type of study and its in subsequent applications it would be useful to observe single-frequency behavior experimentally. There have been recent reports of quantum-cascade lasers operating in the frequencies discussed in this paper,\cite{colombelli01,castellano11} which at present are at the lower frequency limit of this technology in the infrared region.  In the immediate future, spectroscopic measurements, with the aid of stops, slits or gratings, may be an easier option for investigating imaging effects in quartz. However, optical phonons, in general, span a wide frequency range, and investigations of imaging properties using monochromatic radiation may be more straightforward in other materials. The higher frequency modes of calcite,\cite{long93} for example, should be far more easily accessible  using quantum-cascade lasers than those of quartz. At considerably lower frequencies, below $\sim150\mbox{ cm}^{-1}$, quantum-cascade lasers may again be considered as possible sources,\cite{williams07} along with other devices such as backward-wave oscillators.\cite{dobroiu06} Imaging using crystals such as Hg$_2$I$_2$ an TGS, whose phonons fall in this frequency range, should therefore be possible with the aid of such sources. Overall there appear to be ample possibilities for using anisotropic crystals in this way.

\section{\label{sec:conclusions}Conclusions}

In this work we have shown that anisotropic crystals such as quartz should behave as slab lenses around the optic phonon frequencies, even at room temperature. Furthermore, at the frequencies of the TO phonons polarized normal to the surface, subwavelength imaging based on canalization may be possible in sufficiently anisotropic crystals, and we have shown examples of this using quartz. This work clearly needs extending to other frequencies with other materials, and there appear to be various possibilities for experimental studies.

\begin{acknowledgments}
The work was partially financed by the Brazilian research agencies
CNPq and CAPES.
\end{acknowledgments}


%

\printfigures
\end{document}